\documentclass[mathleft]{an}
\usepackage{graphicx}
\usepackage{times}
\usepackage{natbib}

\bibpunct{(}{)}{;}{a}{}{,}

\newcommand{\msun}{\mbox{${\rm M}_{\odot}$}}
\newcommand{\lsun}{\mbox{${\rm L}_{\odot}$}}

\newcommand{\simgt}{\lower.5ex\hbox{$\; \buildrel > \over \sim \;$}}
\newcommand{\simlt}{\lower.5ex\hbox{$\; \buildrel < \over \sim \;$}}
\newcommand{\BV}{Brunt-V\"ais\"al\"a\ }

\def\apj{ApJ}%
\def\apjl{ApJ}%
\def\apjs{ApJS}%
\def\apss{Ap\&SS}%
\def\aap{A\&A}%
\def\azh{AZh}%
\def\mnras{MNRAS}%
\def\solphys{Sol.~Phys.}%
\def\nat{Nature}%
\begin{document}

\Pagespan{789}{}
\Yearpublication{2006}%
\Yearsubmission{2005}%
\Month{11}%
\Volume{999}%
\Issue{88}%

\title{Inference from adiabatic analysis of solar-like oscillations in Red giants}

\author{J. Montalb\'an\inst{1}\fnmsep\thanks{\email{j.montalban@ulg.ac.be}}
\and  A. Miglio\inst{1,2}
\and A. Noels \inst{1}
\and R. Scuflaire \inst{1}
\and P. Ventura \inst{3}
}
\titlerunning{Solar-like oscillations in Red Giants}
\authorrunning{Montalb\'an et al.}
\institute{Institut d'Astrophysique et G\'eophysique de l'Universit\'e de Li\`ege, All\'ee du six Ao\^ut, 17 B-4000 Li\`ege, Belgium
\and 
Charg\'e de Recherches of the Fonds de la Recherche Scientifique, FNRS, rue d’Egmont 5, B-1000 Bruxelles, Belgium
\and
Osservatorio Astronomico di Roma-INAF, via Frascati 33, Monteporzio Catone, Rome, Italy}

\received{30 May 2005}
\accepted{11 Nov 2005}
\publonline{later}


\abstract{The clear detection with CoRoT and KEPLER of radial and non-radial solar-like oscillations in many red giants paves the way to seismic inferences on the structure of such stars. We present an overview of the properties of the adiabatic frequencies and frequency separations of radial and non-radial oscillation modes, highlighting how their detection allows a deeper insight into the properties of the internal structure of red giants. In our study we consider models of red giants in different evolutionary stages, as well as of different masses and chemical composition. We describe how the large and small separations computed with radial modes and with non-radial modes mostly trapped in the envelope depend on the stellar global parameters and evolutionary state, and we compare our theoretical predictions and first KEPLER data.Finally, we find that the properties of dipole modes constitute a promising seismic diagnostic of the evolutionary state of red-giant stars.}

\maketitle
\section{Introduction}

Red giants are cool  stars with an extended convective envelope, which can, as in main sequence solar-like stars,  stochastically excite pressure modes of oscillation.  Although  stochastic  oscillations in a few red giants have already been detected from ground and space observations \citep[e.g.][]{Frandsen02, Joris06, Barban07} it has been only after the photometric  space mission COROT \citep{Baglin02} that an unambiguous detection of radial and non-radial modes in a large number of red-giant stars have been achieved \citep{Joris09, Hekker09, Carrier10}. That confirmation has opened the way to the seismic study of the structure and evolution of these objects that play a fundamental role in fields such as stellar age determination and chemical evolution of galaxies.

About 2000  of the targets observed by CoRoT in the two first  runs of 150 days have been identified as red giants  with solar-like  oscillations in the frequency domain expected from theoretical scalings by \cite{KB95}. Their  spectra show regular patterns  that allowed \cite{Mosser10} to derive  precise values of the large frequency separation. The analysis of the light curve of the sismo-CoRoT target HR 7349  has revealed a rich  solar-like spectrum with 19 identified modes of degrees $\ell=0$, 1 and 2 \citep{Carrier10}.  Moreover,  the first 34 days of science operations of the  KEPLER satellite \citep{kepler09}, have also revealed the presence of solar-like  oscillations in a sample of 50 stars whose frequency at maximum power ($\nu_{\rm max}$) indicates that they are low-luminosity  red-giants \citep{Beddingetal10}. The mean large ($\Delta \nu$) and small frequency separations ($\delta\nu_{02}$),  classically used in the asymptotic interpretation  of solar-like oscillations, were also derived for that sample. All these new and high quality  data, together with  all those we expect in the next years thanks to KEPLER and  CoRoT missions, are the motivation for the present study about the physical interpretation of  the oscillation spectrum of red-giant stars.

In the next sections we present:  first the properties of the stellar models we computed for different sets of fundamental  parameters and input physics; second, the properties of the corresponding adiabatic spectra; and third,  the predicted  values for the frequency separations of acoustic modes, as well as  the inferences  about the  physical properties of red-giant stars that we can extract from these separations .

\section{Stellar Models}

Stellar models were computed with the code ATON3.1 \citep[][and references therein]{Ventura08} for masses between  1.0 and 5.0 \msun\ with a step of 0.5\msun, and  different chemical compositions:  He mass fraction, $Y$=0.25 and 0.278,  and metal mass fraction, $Z$=0.006, 0.01, 0.015, 0.02 and 0.03. The energy transport in the convective regions was modeled with the classic mixing length treatment with  $\alpha_{\rm MLT}=1.6$. For a given chemical composition, models with  $\alpha_{\rm MLT}=1.9$ and  FST treatment of convection \citep{Canutoetal96} were also computed. The evolution of these models were followed from the pre-main sequence until the exhaustion of He in the center for models more massive than 2.5\msun, and until the helium flash for the less massive ones. The core He-burning phase for low-mass ($0.7-2.3$\msun) stars has been also followed starting from  zero age horizontal branch  models for different values of the mass of the He-core (Red clump stars). Microscopic diffusion was not included in these computations but its effects on red giant models \citep{michaud10} are not relevant for the present study.

\begin{figure}
\resizebox{\hsize}{!}{\includegraphics[]{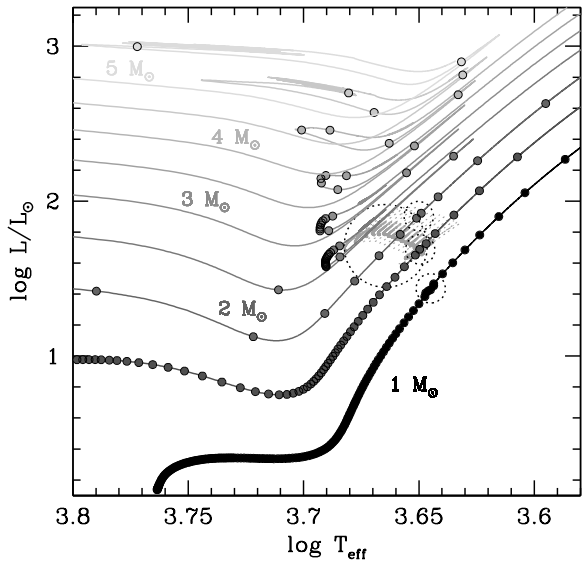}}
\caption{HR diagram for $Z=0.02$, $Y=0.278$, $\alpha_{\rm MLT}=1.9$, and masses between 1 and 5 \msun. Dots on the evolutionary tracks mark a time step of $10^7$~yrs. Small dotted circles indicate the location of the ``Red Bump'' for 1, 1.5 and 2.0 \msun tracks. The large dotted circle surrounds low-mass (0.7 to 2.3 \msun) evolutionary tracks in the phase of post-flash core He-burning (red clump).}
\label{label_HR}
\end{figure}

The rate of evolution during  ascending RGB, descending  RGB and core He-burning phases is very different and strongly  depends on the stellar mass. For  low-mass stars the  time scale  in the RGB  may be comparable with that of  core He-burning of more massive stars (see Fig.~\ref{label_HR}). As a consequence, observing stars in both evolutionary stages would be equally likely. Concerning the internal structure of these  models, it is worth mentioning that  for a low-mass model (1.5~\msun, for instance) the density contrast ($\rho_c/\langle\rho\rangle$, central to mean density ratio) changes from $10^6$  at the bottom of its RGB to  $3\times 10^9$  at $\log L/L_\odot\sim 2$. Models in the core He-burning phase, on the other hand,  have a value of  $\rho_c/\langle\rho\rangle$ of the order of $2\times 10^7$, and due to  the high dependence on temperature of the 3$\alpha$ nuclear reactions, have developed  a small convective core. At a given  luminosity, that of the Red-Clump for instance,  $\rho_c/\langle\rho\rangle$ for a 1.5\msun\ RGB model is more than 10 times larger than for  He-burning one. So different structures should imply significant effects on the oscillation properties.

\begin{figure*}[ht!]
\centering
\resizebox{\hsize}{!}{\includegraphics{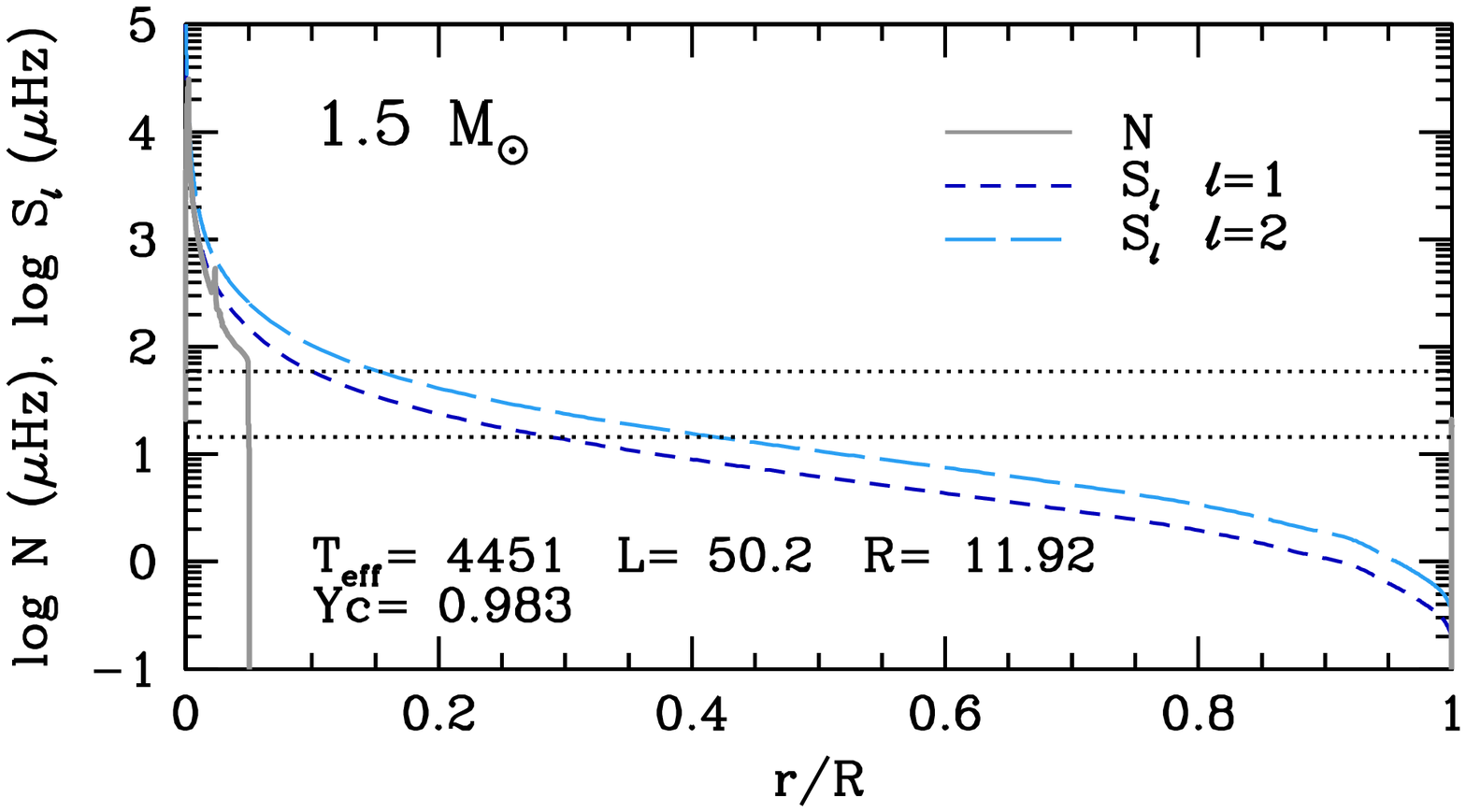}\includegraphics{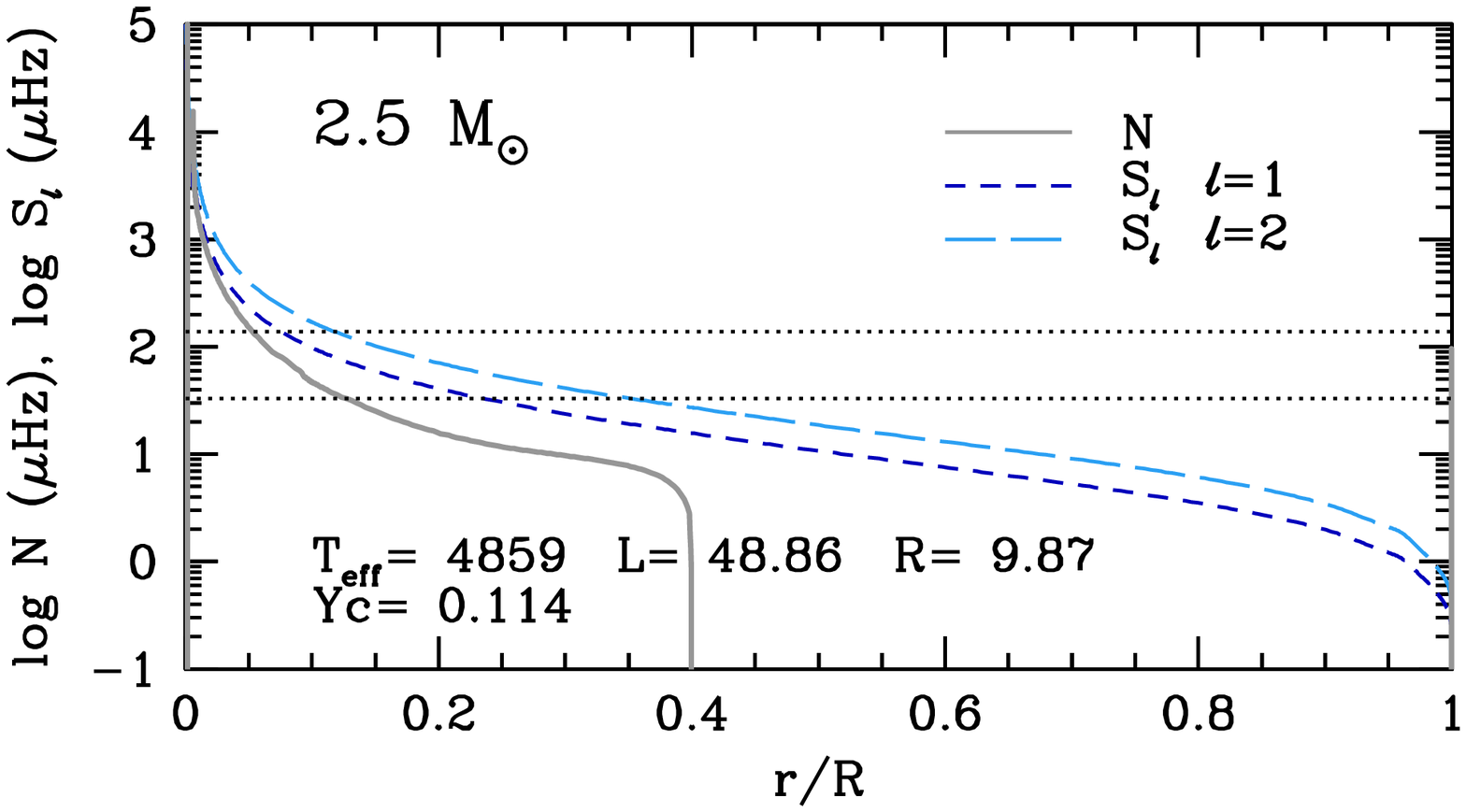}}
\resizebox{\hsize}{!}{\includegraphics{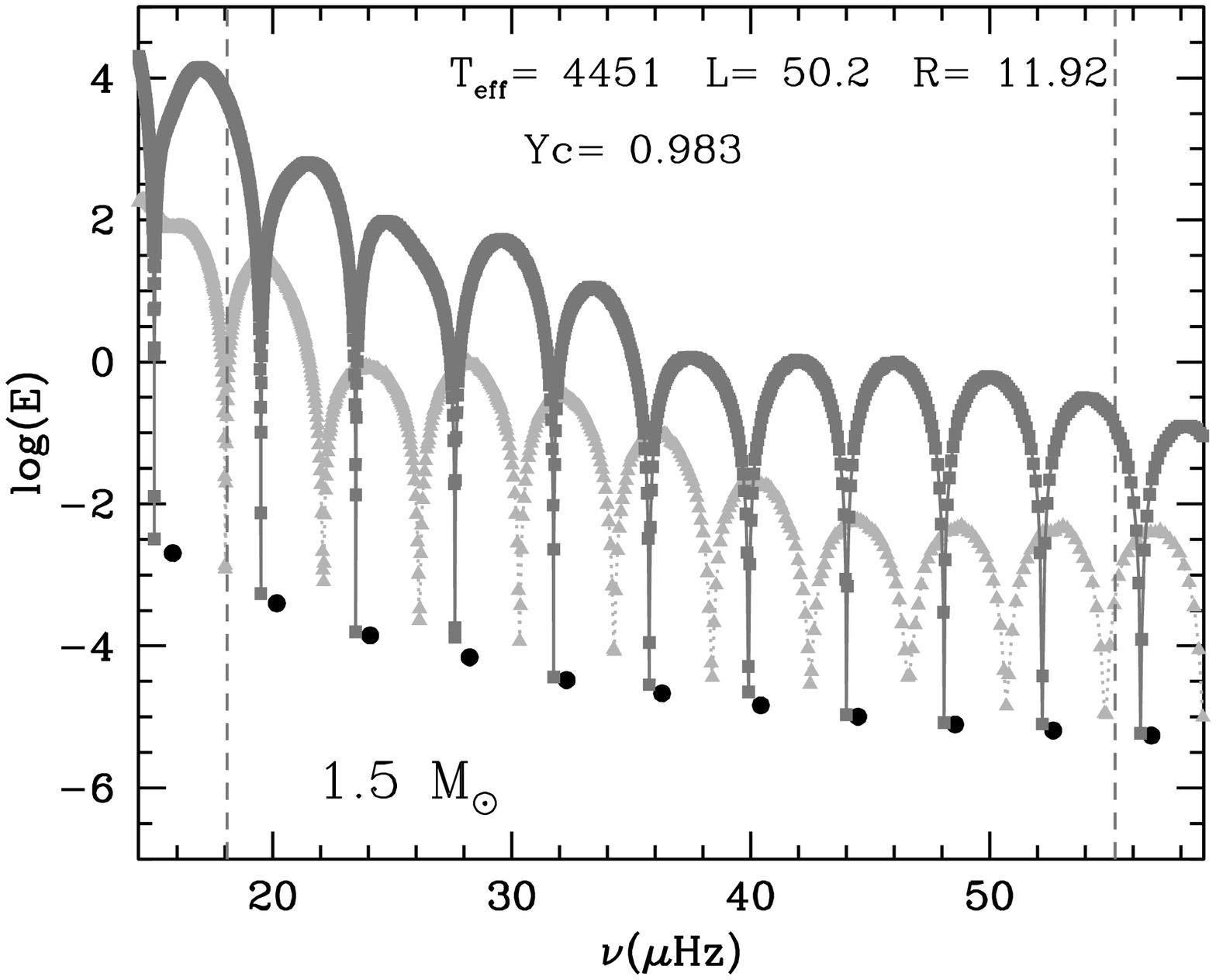}\includegraphics{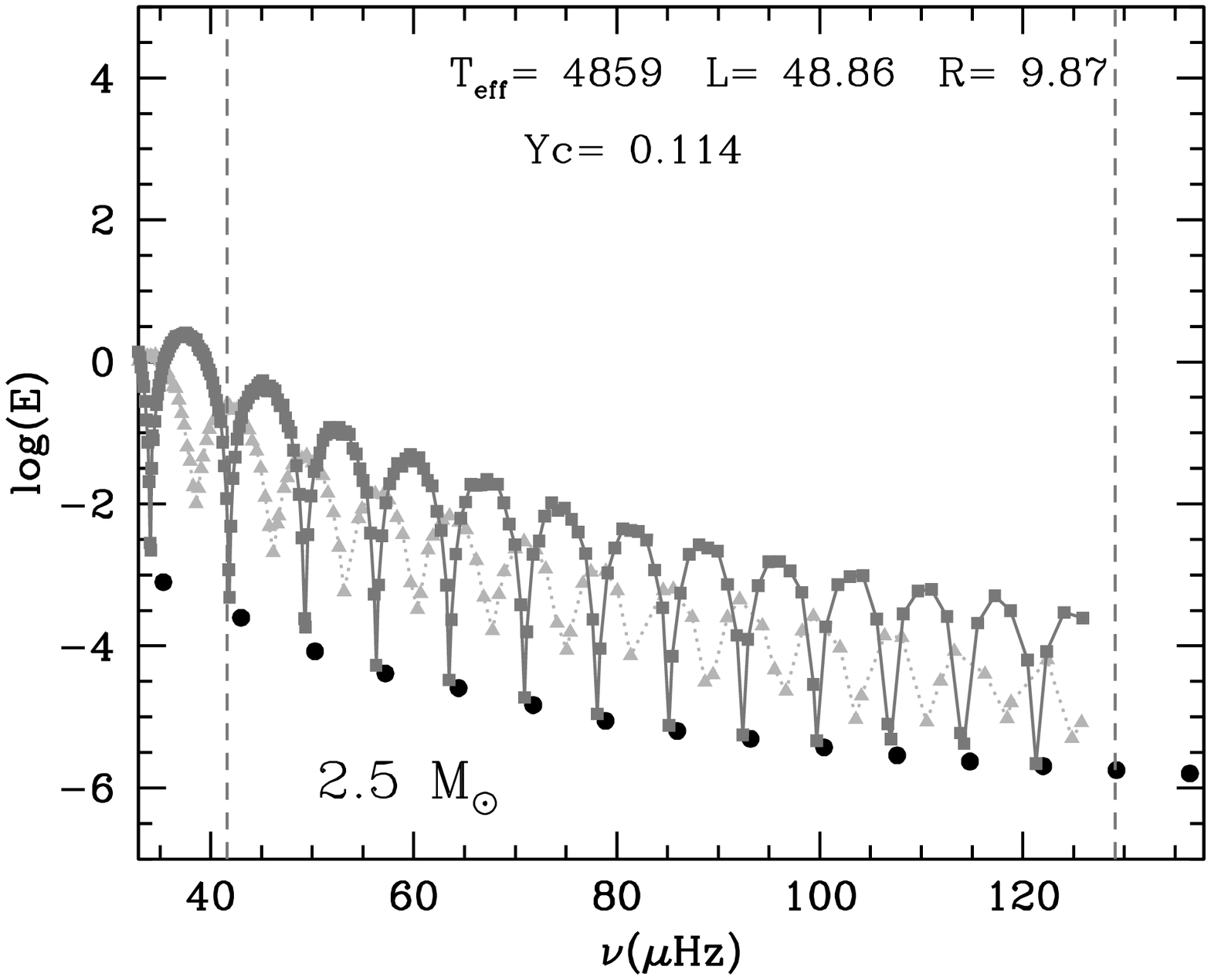}}
\vspace*{-1.75cm}
\caption{Upper panels: propagation diagrams for two models at almost the same luminosity,  1.5~\msun\ model  in the RGB (left) and 2.5\msun\ model in central He-burning phase (right). Horizontal dotted lines limit the solar-like frequency domain for each model. Solid line is \BV\ frequency, and short- and long-dashed lines correspond to the Lamb frequency for $\ell=1$ and 2 respectively.  Lower panels: corresponding plots of  inertia as a function of frequency for $\ell$=0 (black circles), 1 (grey triangles) and 2 (dark-grey squares) modes in the solar-like oscillation domain delimited by dashed vertical lines.}
\label{label_prop}
\end{figure*}

\section {Adiabatic oscillation properties}

Adiabatic oscillation frequencies were computed with an Eulerian version of the code LOSC \citep{Scuflaire08} for models from the bottom of the RGB until a maximum luminosity  ($\log L/L_\odot ~ 2.2-3.2$, depending on mass) and also during the phase  of core  He-burning.
In this paper we deal with adiabatic computations and we do not consider the problem of excitation and damping of solar-like oscillations in red giants \cite[][]{Dziembowski01, HoudekGough02, Dupret09}. We use the scaling laws \citep[][]{Brownetal91, KB95} to derive  the frequency domain in which solar like oscillations are expected, and the value of the mode inertia as an estimate of the expected mode amplitude \citep[][]{JCD04}.  We search oscillation modes with angular degree $\ell=0$, 1 and 2, in the domain of frequencies defined by  an interval around $\nu_{\rm max}$ \citep[Eq. (10) of][]{KB95}. The width of the solar-like frequency domain is taken to be 20\% larger than the  difference between the acoustic cutoff frequency in the stellar atmosphere  and $\nu_{\rm max}$.

The  properties of oscillation modes depend on the behaviour of the \BV\ ($N$) and Lamb ($S_{\ell}$) frequencies. In red-giant models, $N$ reaches large values in the central regions and therefore the frequency of gravity modes (g~modes)  and the number of g-modes by frequency interval ($n_g$) increase with respect to main sequence models. The value of $n_g$ can be estimated from the asymptotic theory (Tassoul 1980) as:
\begin{equation}
n_g \propto [\ell\,(\ell+1)]^{1/2}\int \frac{N}{r} dr,
\label{eq1}
\end{equation}
\noindent with the integral performed on the radiative region.
On the other hand, the low mean density makes  the frequency of pressure modes (p-modes) to decrease. All that leads to an oscillation spectrum for red-giants where in addition to   radial modes, there is a large number of non-radial modes with mixed g-p properties. The dominant character of these non-radial modes depends on the separation between gravity and acoustic cavities, and may be estimated from the value of the normalized  mode inertia ($E$) \citep[see e.g.][ and references therein]{JCD04}. Therefore, some non-radial modes may be well trapped in the acoustic cavity and behave as p-modes presenting a  mode inertia close to that of radial modes, while modes with strong mixed g-p character have larger $E$ value.  Hereafter, we will use the term p-modes in quotation marks to refer to mixed modes with a dominant p-character.

 In Fig.~\ref{label_prop} we present, in top panels, the $\ell=1$, 2  propagation diagrams for a RGB 1.5~\msun\ model (left) and for a core He-burning (He-B) model of 2.5\msun\ (right).  In the bottom panels we plot the variation of the mode inertia with frequency for radial and non-radial  modes ($\ell=1,\,2$). The RGB model is ten times more centrally condensed than the He-B one ($\rho_c/\langle \rho\rangle \sim 5\,10^8$ and $10^7$, respectively). Furthermore, the He-B model has a small convective core (too small to be seen at the scale of Fig.~\ref{label_prop}). Both ingredients contribute to decrease the value of $N$ in the central regions and therefore  the density of non-radial modes (Eq.~1). While the number of $\ell=1$  eigenfrequencies between two radial ones is of the order of 200 for the RGB model, it is only  6 for the He-B one.
The huge difference in  density between the central region and the convective envelope entails a high potential barrier between the acoustic and the gravity cavities reducing the interaction between p and g modes. As a consequence,  we find   $\ell=1$ modes  with $E_{\ell=1}\approx E_{\ell=0}$ that are quite regularly spaced in frequency for RGB models. For He-B ones,  the coupling between these cavities is more important and $\ell=1$ modes are mixed modes with $E_{\ell=1}> E_{\ell=0}$. Nevertheless, $E_{\ell=1}$ presents  still a minimum value for modes between two consecutive radial ones showing a  somewhat regular pattern. Even if $E$ is larger than that corresponding to the radial modes  we can consider those modes, based on the value of $E$, as still observable ``p-modes''. For $\ell=2$ modes, the interaction  between g- and p-cavity is smaller  than for $\ell=1$ and hence the trapping more efficient.  Therefore, independently of the central  condensation of the model,  a spectrum of regularly spaced $\ell=2$ ``p-modes''  with $E_{\ell=2}\approx E_{\ell=0}$ is expected.
Finally, note that  the turning points for acoustic modes ($tp_\ell$ defined as point where $\nu_{\rm max}=S_{\ell}$) are inside the convective envelope for RGB model and in the radiative region for the He-B one.

In the asymptotic theory for p-modes \citep{vandakurov67, tassoul80, gough86} the frequency of two consecutive modes of same degree are separated by a constant value $\langle\Delta \nu\rangle$   and that value is  approximately  independent of $\ell$. Of course, the asymptotic theory is no longer valid for mixed modes nor in regions with rapid variation of physical quantities, but we can, nevertheless, try to apply it to the modes  well or partially  trapped in  the  acoustic cavity, with a dominant p-character (``p-modes''). We select then  as ``p-modes'' of degree $\ell$ the modes with  the  minimum inertia  between  two consecutive radial modes, and we use them to derive the values of the large and small frequency separations and to analyse the dependence of these asymptotic quantities on the stellar parameters and evolutionary state.

\subsection{Large Separation: $\Delta\nu$}

\begin{figure*}[ht!]
\centering
\resizebox{\hsize}{!}{\includegraphics{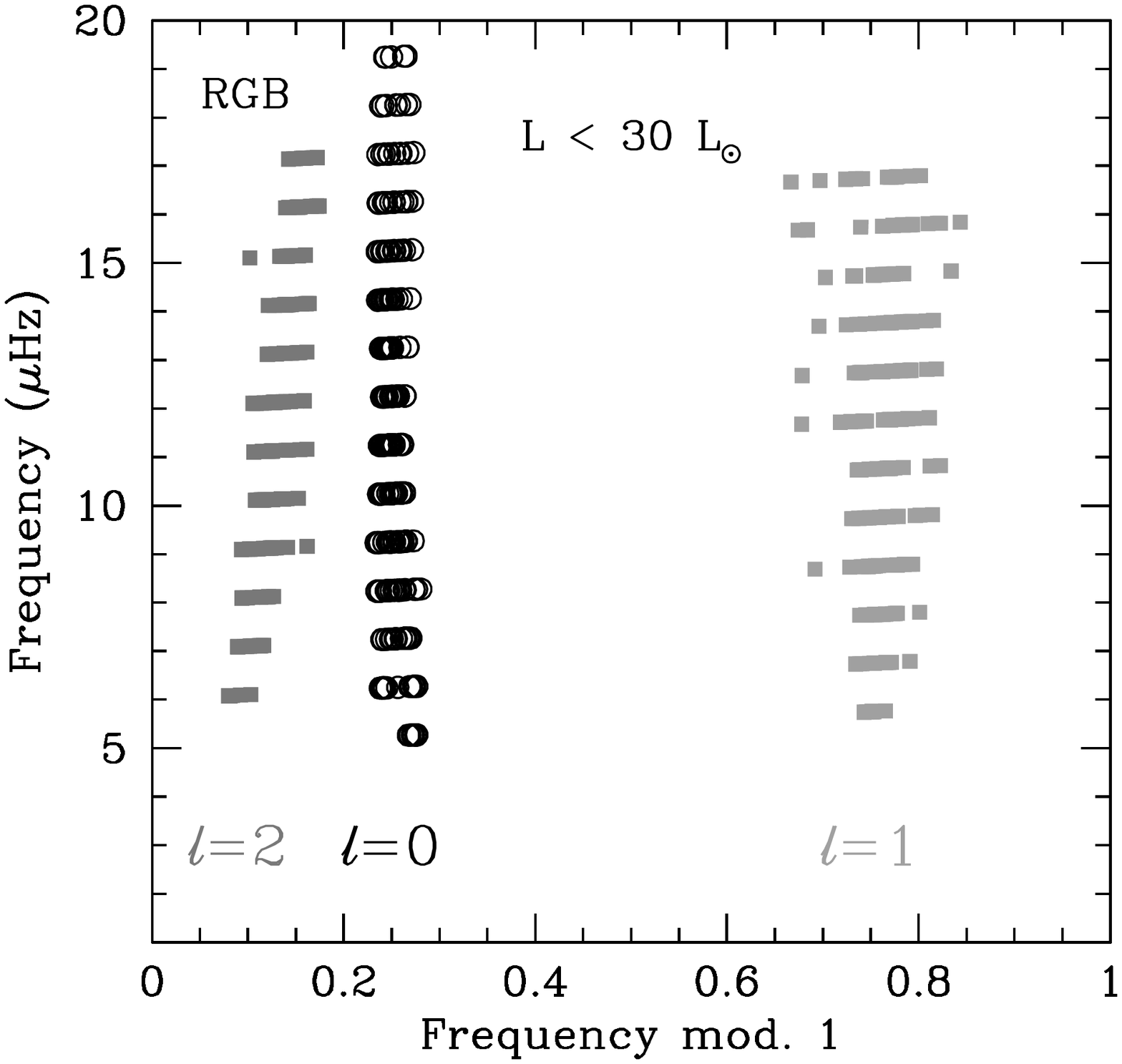}\includegraphics{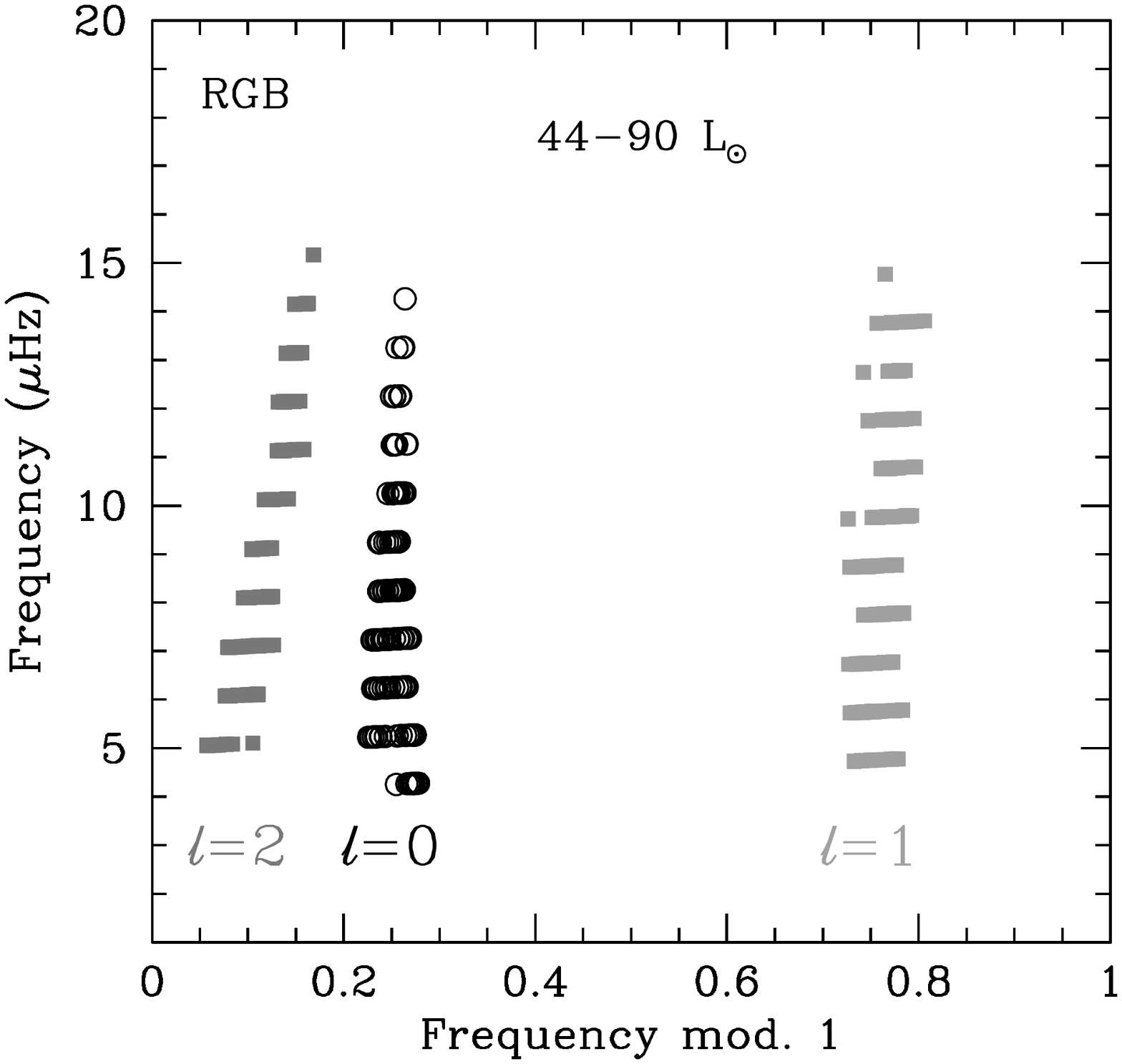}\includegraphics{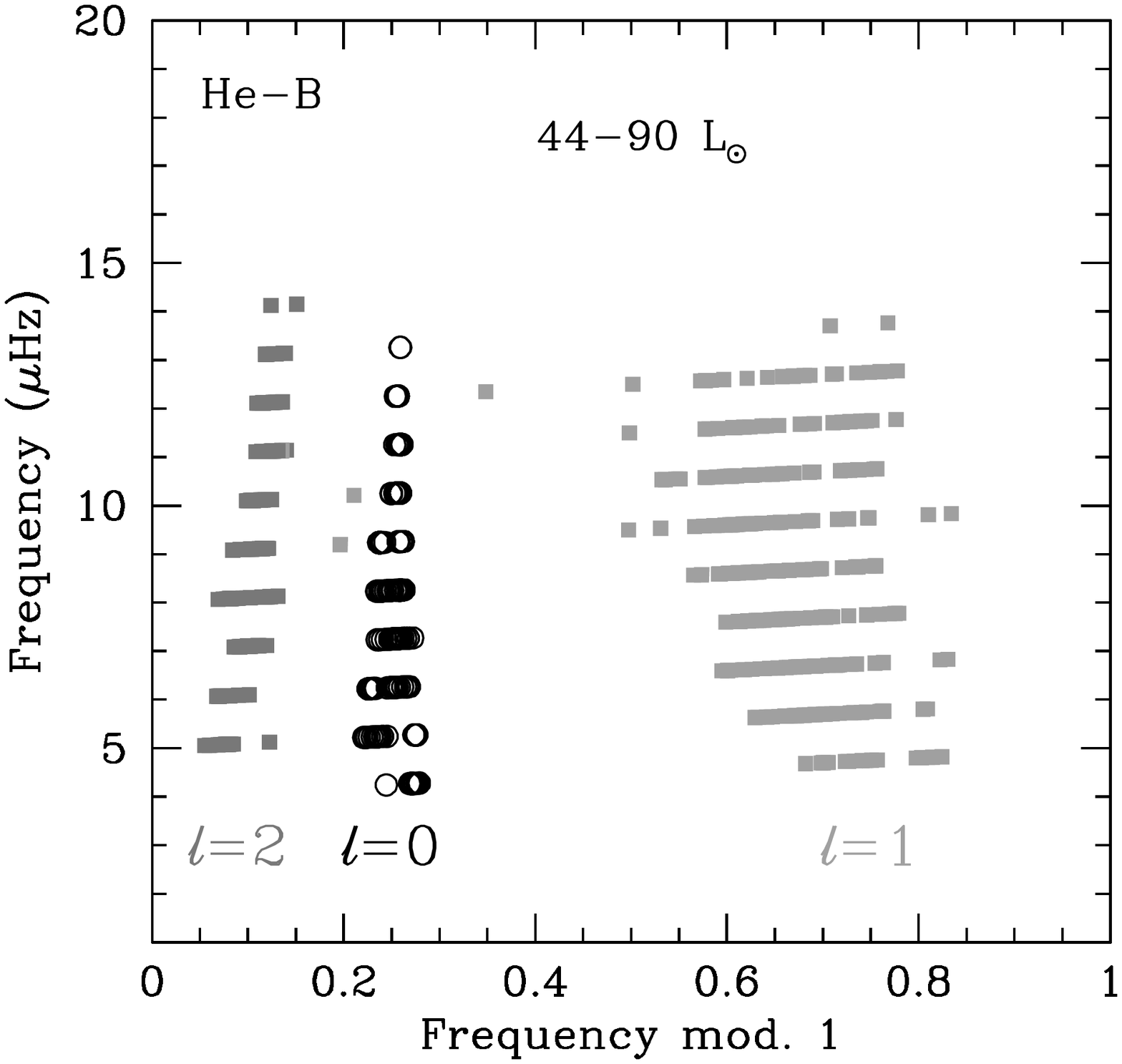}}
\caption{Folded \'echelle diagram \citep[see i.e.][]{BK10} for 1 and 1.5\msun\ models in different evolutionary stages:  RGB with luminosity lower than 30\lsun\ (as assumed in the first red giants observed by KEPLER) in the left panel; RGB models with luminosity  within the Red Clump domain in the central panel; He-burning models with luminosity between 44 and 90 \lsun in the right panel .}
\label{label_echell}
\end{figure*}

The mean value of the large frequency separation ($\langle\Delta\nu_{\ell}\rangle=\langle\nu_{n,\ell}-\nu_{n-1,\ell}\rangle$) averaged over the radial order $n$  is, according to the asymptotic theory,  related to the mean density of the star, and its value is approximately  independent of $\ell$ for low-degree modes. Therefore, for a given mass, $\langle\Delta\nu\rangle$ decreases as the stars ascends in the RGB with a core more and more dense and an envelope more and more diffuse, but it is not possible  to distinguish, on the base of $\langle\Delta\nu\rangle$  value, among different evolutionary states, ascending RGB, descending RGB, or core He-burning.   Nevertheless, there is an indirect information about the evolutionary state not in $\langle\Delta\nu\rangle$, but on the deviation of $\Delta\nu$ as function of frequency with respect to its mean value ($\sigma(\Delta\nu_{\ell})$). Radial and  $\ell=2$ ``p-modes'', as  mentioned above, show a very regular pattern, and  the mean quadratic deviation of $\Delta\nu(\nu)$ with respect to its mean value over the solar-like frequency domain ($\sigma(\Delta\nu_{\ell})$) is always below 5\% for all the evolutionary states and masses considered.  On the contrary, $\sigma(\Delta\nu_1)$ strongly depends on the evolutionary state and  while its value remains small for more concentrated models, it may get values as large as 30\% for core He-burning models.

\begin{figure*}[ht!]
\centering
\resizebox{\hsize}{!}{\includegraphics{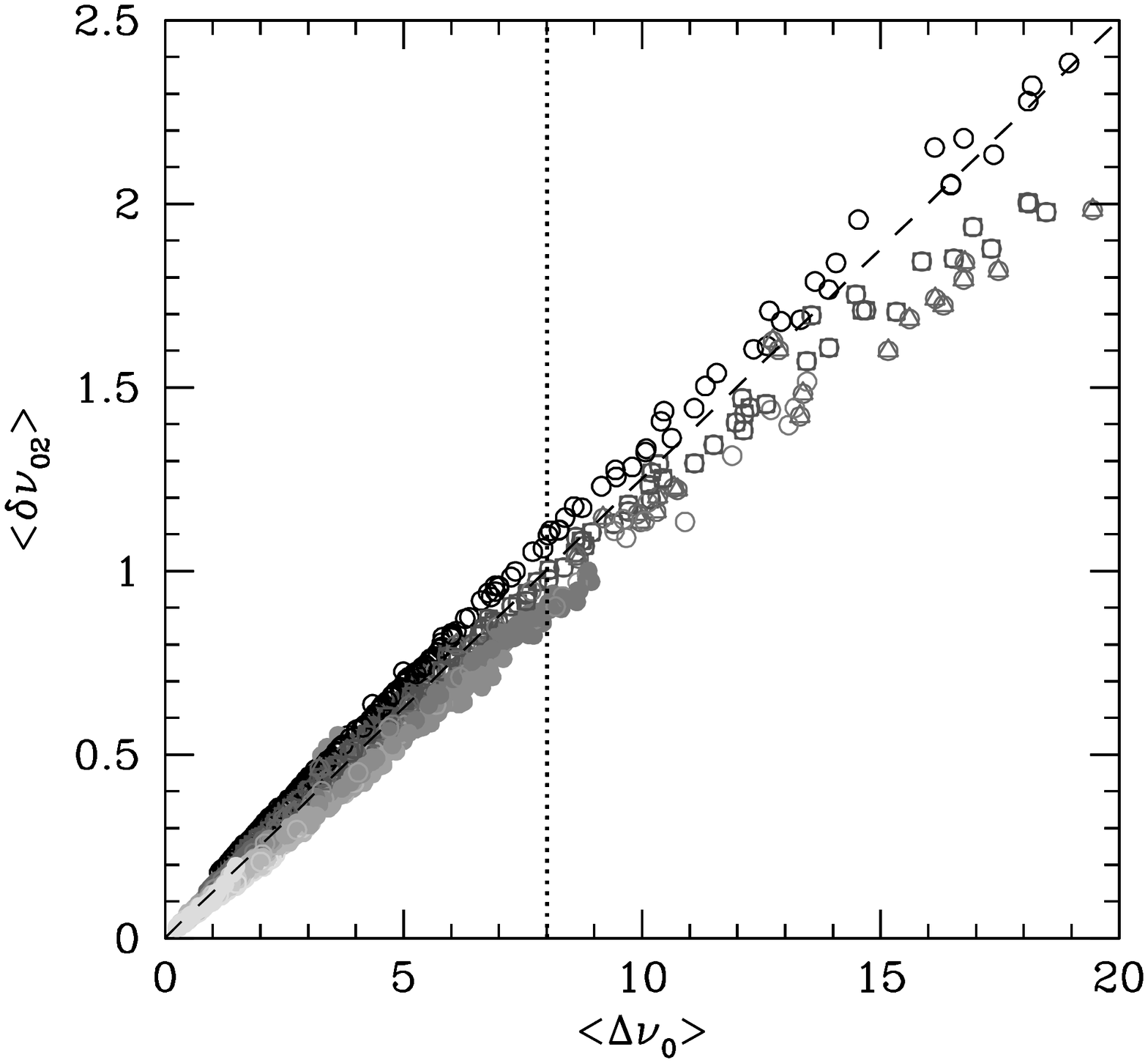}\includegraphics{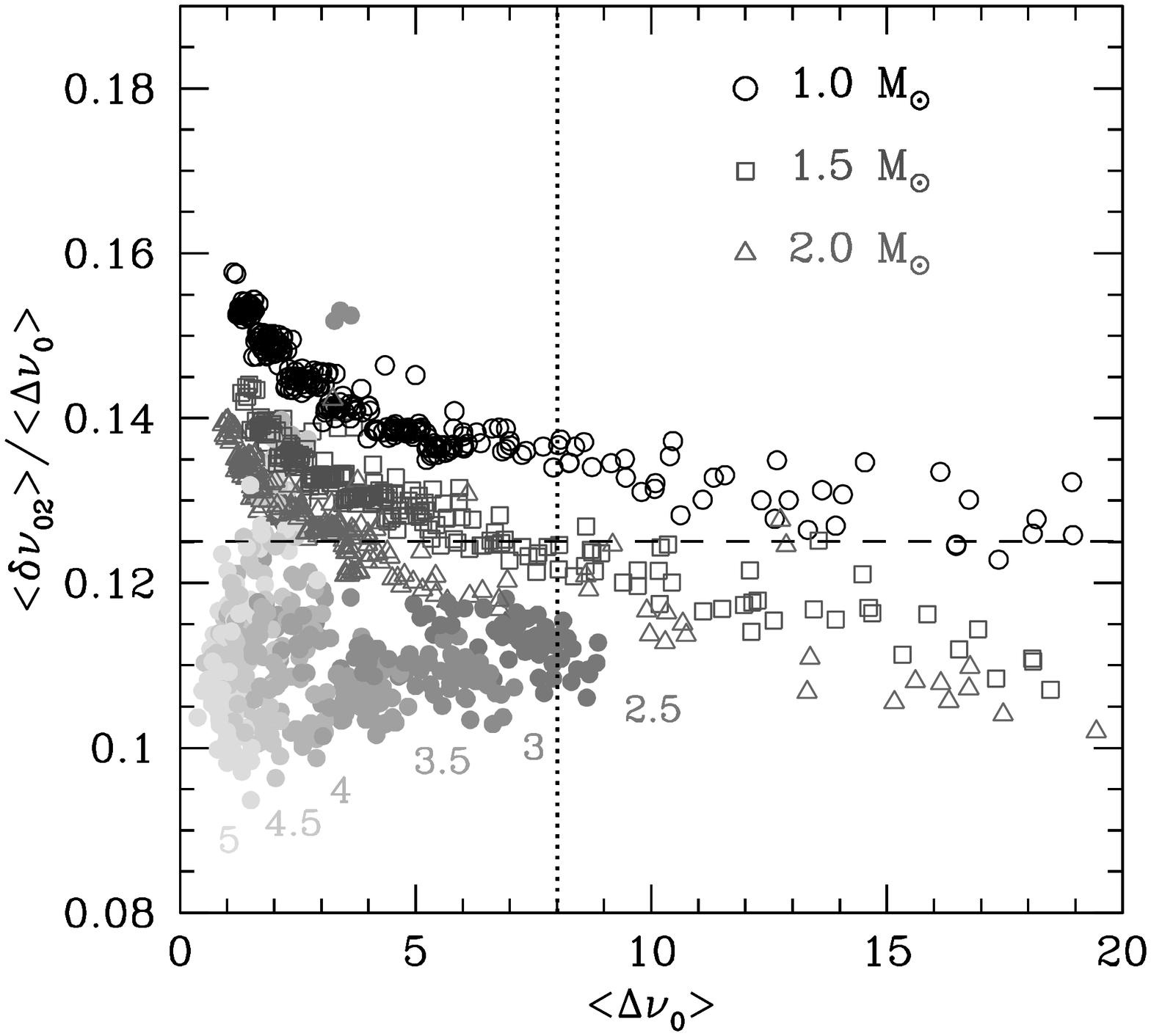}\includegraphics{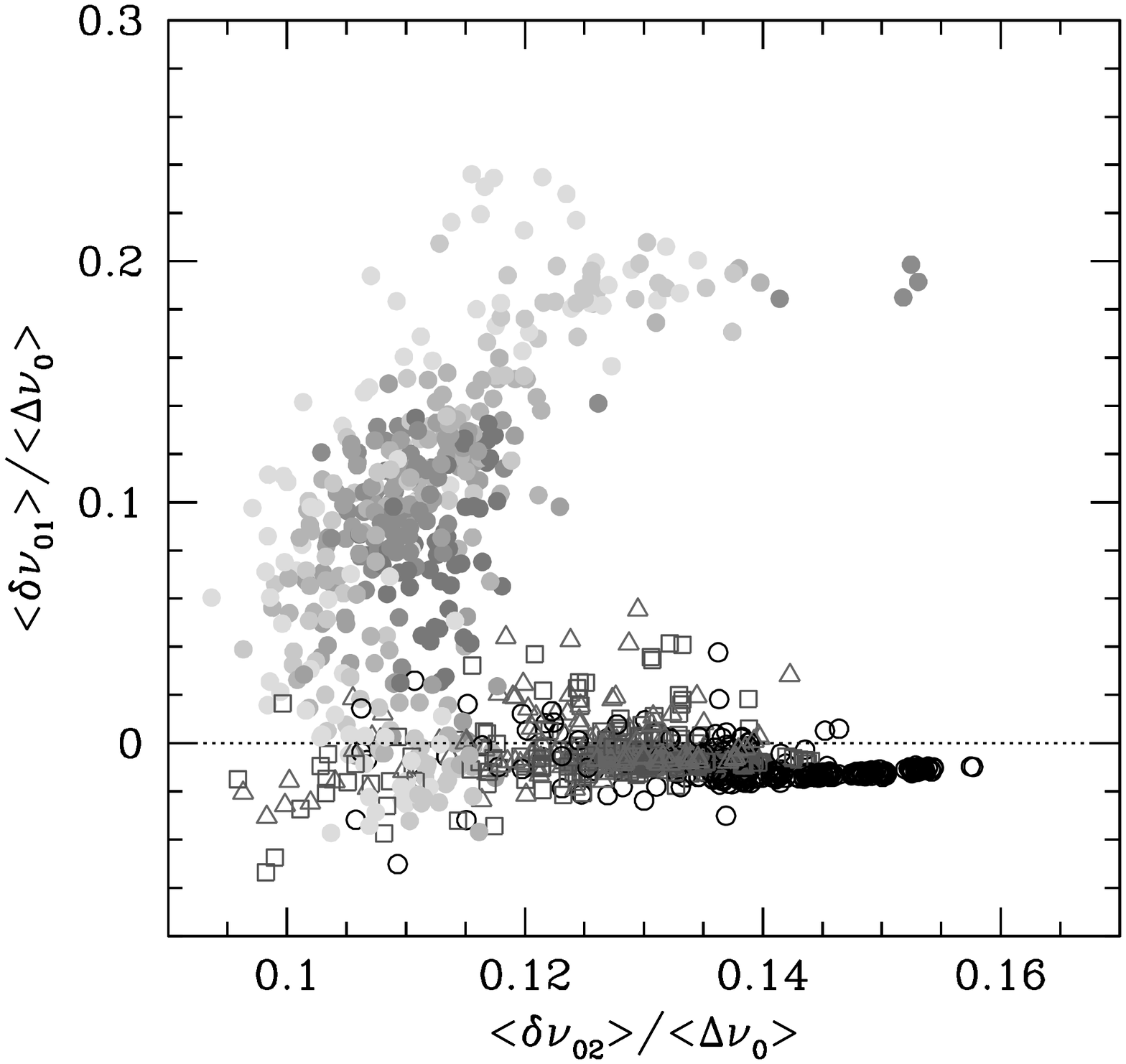}}
\caption{a) Mean value of the small separation frequency $\delta\nu_{02}$ as a function of the mean value of  large frequency separation for radial modes. Open symbols correspond to low-mass models (1, 1.5 and 2. \msun) from the bottom of the RGB up to luminosity $\log L/L_\odot ~ 2.5--3$, solid grey dots correspond to models in the He-burning phase ($Y_c$ between 0.9 and 0.1) with masses between 2.5 (darkest grey dots) and 5\msun\ (light grey dots). These results correspond to models with $Z$=0.006, 0.015, 0.02 and 0.03, $Y$=0.025 and 0.278, $\alpha_{\rm MLT}=1.6$ and 1.9, and models with and without core overshooting. Dotted vertical lines limits the domain of large separation obtained in the first sample of red giants observed by  KEPLER, and the dashed line correspond to the linear fit for these data from  \cite{Beddingetal10}. b) Same as a) but for the small separation frequency $\delta\nu_{02}$ normalized to the large separation. c) Normalized small frequency separation $\delta\nu_{01}$ as a function of the normalized small separation $\delta\nu_{02}$ for the same models as in a) and  b) panels. Note the high concentration of RGB models with small and negative value of $\delta\nu_{01}$.}
\label{label_d02}
\end{figure*}

Following the Method 2 in \cite{BK10},  we plot the adiabatic frequencies of trapped ``p-modes''  for models with 1 and 1.5 \msun\ in the form of folded \'echelle diagram. The left panel in Fig.~\ref{label_echell} presents the spectrum for models with luminosity between the bottom of the RGB and 30\lsun. As found in the observational results obtained by \cite{Beddingetal10} for the first 34 days of KEPLER observations, radial and quadruple modes show a low scatter, while for  dipole modes that one  is significantly larger. As the star goes up in the RGB,  the $\ell=1$ modes are better trapped in the acoustic cavity and the spectra of dipole modes are more regular. That is shown in the central panel of Fig.~\ref{label_echell} where the folded \'echelle diagram for  RGB models at the luminosity of the Red Clump show a smaller scatter for $\ell=1$ modes than in left panel.  At the same luminosity, models with 1 and 1.5 \msun\ may be burning He in their core. That implies a decrease of the central density as well as the development of a small convective core. 
Furthermore, following the expansion of the core, the external convective zone recedes increasing the interaction between the gravity and acoustic cavities (see. Fig.~\ref{label_prop}) and the mixed character of oscillation modes. The spectrum of ``trapped'' $\ell=1$ modes is then much less regular as is made evident in the \'echelle diagram in the right panel of Fig.~\ref{label_echell}. 

\begin{figure*}
\centering
\resizebox{\hsize}{!}{\includegraphics{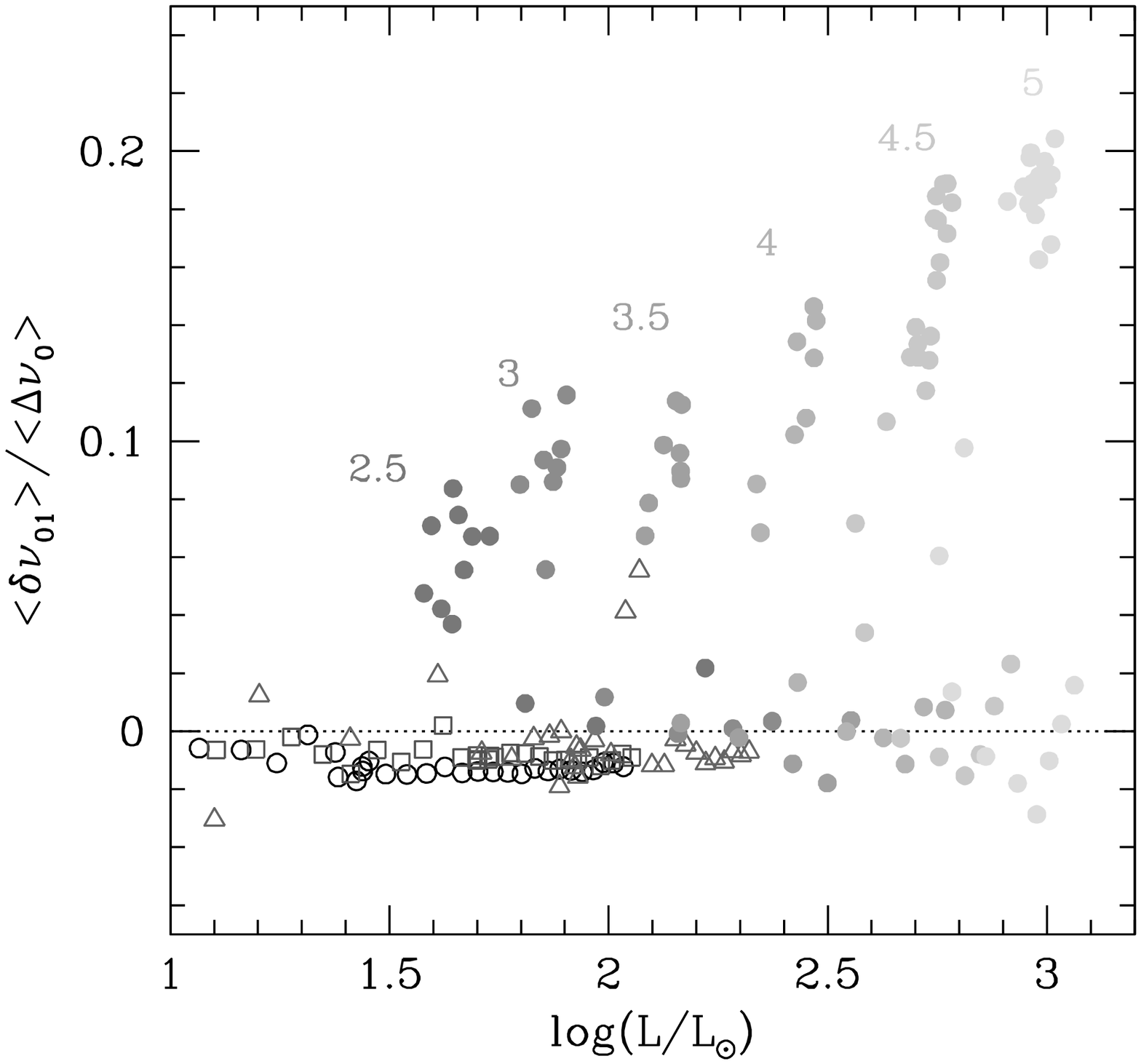}\includegraphics{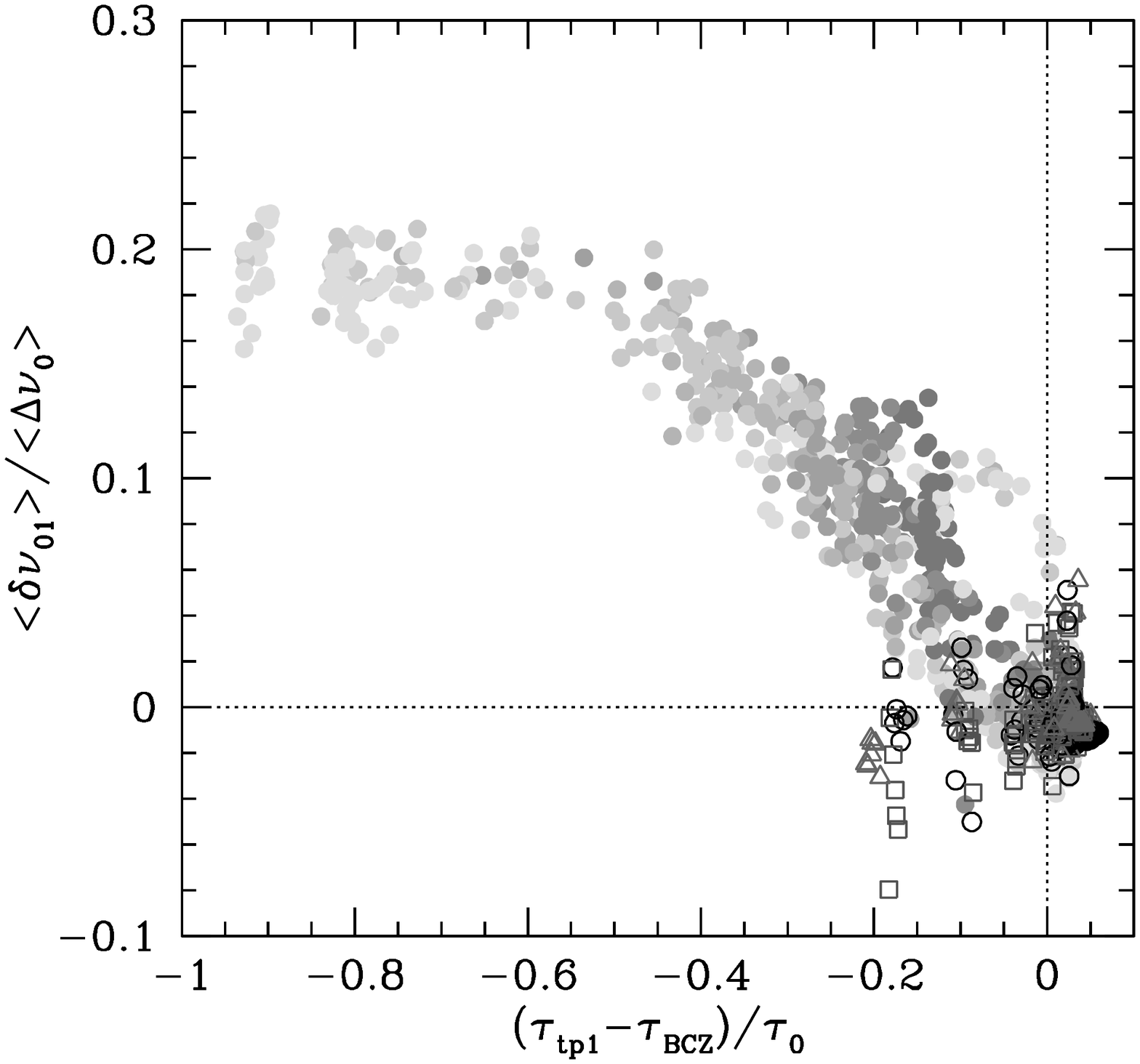}\includegraphics{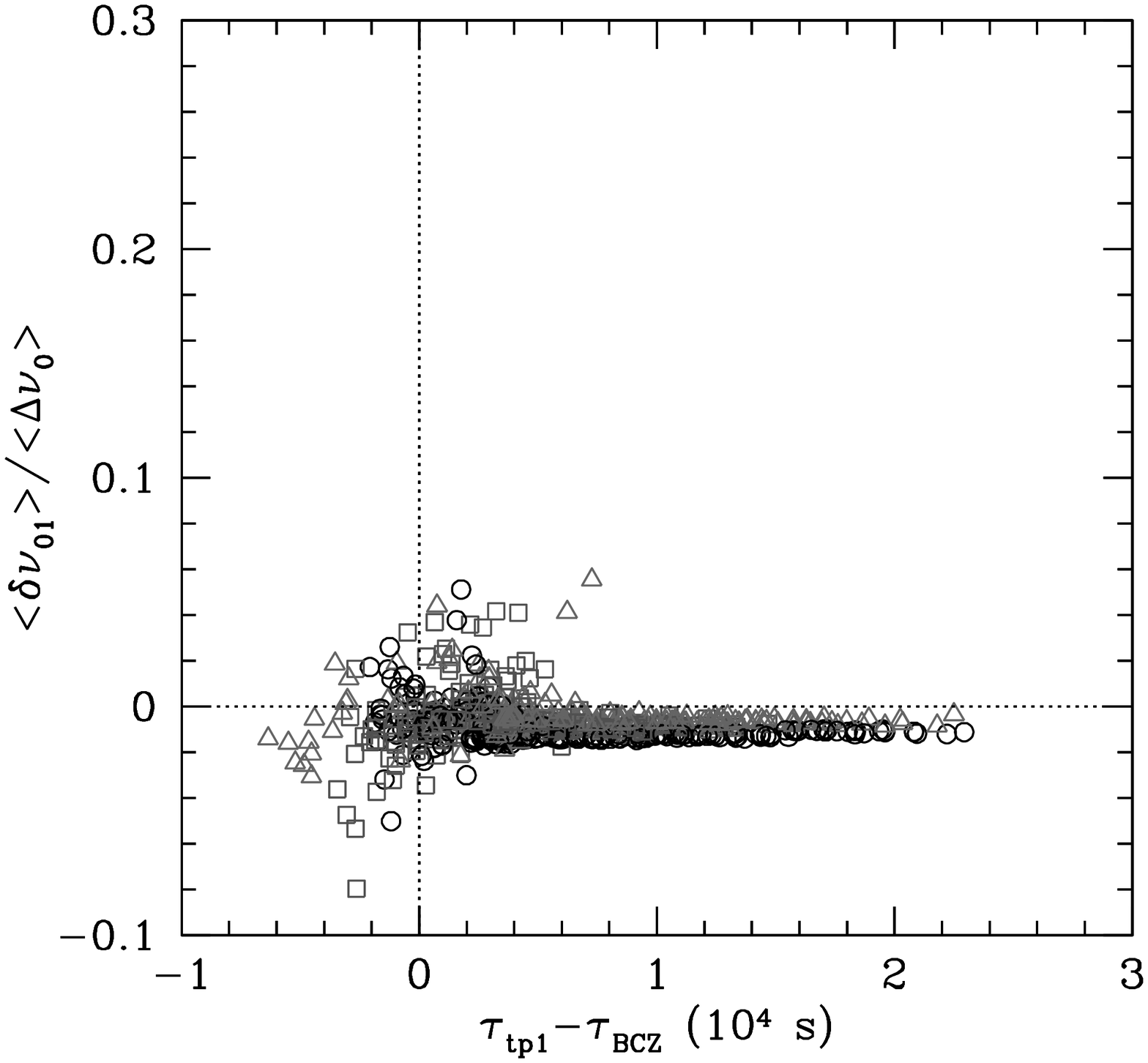}}
\caption{a) Normalized small separation $\delta\nu_{01}$ as a function of luminosity for models with initial chemical composition $Z=0.02$, $Y=0.278$, and $\alpha_{\rm MLT}=1.9$. Open symbols corresponds to 1, 1.5 and 2. \msun\ models and solid grey dots to models with masses between 2.5 and 5\msun. b) Normalized $\delta\nu_{01}$ small separation as a function of the difference in relative acoustic radius between the bottom of the convective envelope $\tau_{\rm BCZ}/\tau_0$  and the $\ell=1$ turning point $\tau_{\rm tp1}/\tau_0$. c) Normalized $\delta\nu_{01}$ small separation as a function of  distance in seconds between the bottom of the convective envelope and the turning point of $\ell=1$ modes, for models with masses between 1 and 2.\msun in the RGB and the same physical inputs as in b) panel.  }
\label{label_d01L}
\end{figure*}

\subsection{Small separations: $\delta\nu_{02}$ and $\delta\nu_{01}$}

The  mean small frequency separation ($\langle\delta\nu_{02}\rangle=\langle\nu_{n0}-\nu_{n-1\,2}\rangle$) is, according to the asymptotic theory, related to the behaviour of the sound speed ($c$) mostly in the central regions, and hence to the stellar evolutionary state. The representation of $\langle\delta\nu_{02}\rangle$ vs. $\langle\Delta\nu\rangle$ is in fact considered as a seismic diagnostic diagram allowing to derive stellar mass and age for main-sequence solar-like pulsators \citep{jcd88}. The corresponding seismic diagram for red-giant models is drawn in Fig.~\ref{label_d02}.a  which shows a linear  dependence of $\langle\delta\nu_{02}\rangle$ on $\langle\Delta\nu\rangle$, with a slope that increases as mass decreases  for RGB models. In Fig.~\ref{label_d02}.b we plot the normalized quantity  $\langle\delta\nu_{02}\rangle/\langle\Delta\nu\rangle$ that in main-sequence stars is known to depend mostly on  central physical conditions \citep{RoxburghVorontsov03}.  It is worth noticing that in these figures we plotted  $\langle\delta\nu_{02}\rangle$ for models computed with different chemical compositions and convection treatment and, nevertheless, a predominant  dependence on mass and radius appears. For a given mass  $\langle\delta\nu_{02}\rangle/\langle\Delta\nu\rangle$ increases with density contrast, its value decreases as mass increases and it does not change significantly during the core He-burning phase.

In Fig.~\ref{label_d02}.a and b  a vertical dotted line indicates the lower limit of $\langle\Delta\nu\rangle$ measured from the first 34d of KEPLER mission,  and  the dashed-line corresponds to the fit  $\langle\delta\nu_{02}\rangle=0.122\langle\Delta\nu_0\rangle$ proposed by \cite{Beddingetal10} for those observations.  The comparison between these figures and  Fig.4 in \cite{Beddingetal10} indicates that predictions from  theoretical models are consistent with  observational data corresponding to low-mass stars (1-1.5\msun) in the low-luminosity part of the ascending RGB, such as the scaling based on $\nu_{\rm max}$ and $\Delta\nu$ also suggests.

The small frequency separation $\delta\nu_{01}(n)=0.5\,(\nu_{0n}-2\,\nu_{1,n}+\nu_{0\,n+1})$ in main-sequence stars is also known  to be  sensitive to the center physical conditions and it is mainly useful when only radial and dipole modes are observed. The asymptotic theory predicts a $\langle\delta\nu_{01}\rangle=1/3\langle\delta\nu_{02}\rangle$ relationship. As it is evident in  Fig.~\ref{label_d02}.c  the ``p-mode'' spectrum for red-giant models does not follow those predictions. In particular, a large number of models have negative or very small values of $\langle\delta\nu_{01}\rangle$  independently of $\langle\delta\nu_{02}\rangle$.  Very small or negative values of $\langle\delta\nu_{01}\rangle$ have also been observed in the KEPLER data \citep{Beddingetal10} and in the oscillation spectrum of the CoRoT red-giant HR~7349 \citep{Carrier10}.  We note that the largest concentration of negative/small $\langle\delta\nu_{01}\rangle$ values   correspond to low-mass models in the RGB. A plot of $\langle\delta\nu_{01}\rangle/\langle\Delta\nu\rangle$ vs. luminosity (Fig.~\ref{label_d01L}.a) suggest in fact that models with negative or small $\langle\delta\nu_{01}\rangle$ value are in or close to the RGB. 

Searching for a common characteristic in the structure of these models, we find that during the ascending and descending RGB the turning points of $\ell=1$ modes are well inside the convective envelop. The steady He-burning models have a shallower convective envelope and the turning points of $\ell=1$ modes are inside the radiative region. Fig.~\ref{label_d01L}.b shows the variation of  $\langle\delta\nu_{01}\rangle/\langle\Delta\nu\rangle$  with the distance (in relative acoustic radius $\tau(r')=\int_0^{r'} dr/c$) between the bottom of the convective zone (BCZ) and the turning point for a $\ell=1$ mode with frequency close to $\nu_{\rm max}$ ($tp_1$). In Fig.~\ref{label_d01L}.c we highlight the behaviour of $\langle\delta\nu_{01}\rangle$ for low-mass models in the RGB as a function of the difference (in seconds) between the acoustic radius of $tp_1$ and that of the BCZ.  We see that the scatter of $\langle\delta\nu_{01}\rangle/\langle\Delta\nu\rangle$ rapidly decreases as $\tau_{tpl_1}-\tau_{\rm BCZ}$ increases (deep convective envelope) and  $\langle\delta\nu_{01}\rangle/\langle\Delta\nu\rangle$ takes negative values for models in which the $tp_1$ is well inside the convective envelope.

\section{Concluding remarks}

In this paper we presented the properties of the oscillation spectrum of solar-like oscillations during the  RGB and core He-burning phases of red-giant evolution, and analyzed the behaviour of large and small frequency separations derived for modes well trapped in the acoustic cavity of these stars. The main results of this global overview are the followings:

\begin{itemize}
\item Independently of the evolutionary state, $\ell=2$ modes trapped in the acoustic cavity have an inertia of the same order as that of the corresponding radial mode and behave as ``p-modes'' with frequencies regularly spaced by $<\Delta\nu>$. As a consequence, the scatter of $\ell=2$ modes in the folded \'echelle diagrams is rather small.

\item The trapping of $\ell=1$ modes in the acoustic cavity depends on the evolutionary state. While a regular pattern of dipole modes is expected in more centrally  condensed models, the scatter significantly increases for models in the core He-burning phase. The scatter of $\ell=1$ ``p-modes'' decreases as model concentration increases. Therefore the regularity  of $\ell=1$ spectrum  could be used to discriminate between different evolutionary phases.

\item $\langle\delta\nu_{02}\rangle$  depends almost linearly on the large separation, hence on the mean density of the model, with a slope that slightly depends on the mass. The value of $\langle\delta\nu_{02}\rangle/\langle\Delta\nu_0\rangle$  for a given mass increases with the density contrast and thus with luminosity, and  for a given luminosity, it decreases as the mass of the model increases.

\item $\langle\delta\nu_{01}\rangle$ seems to  reflect the distance between the $\ell=1$ turning point and the bottom of convective envelope. It takes negative (or small) values if tp1 is well inside the convective envelope, what occurs in RGB models.

\item  The theoretical predictions based on stellar models are in good agreement with the observational results obtained in the first 34 days of KEPLER observations. Comparison of their $\nu_{\rm max}$  values with  the population simulations presented in \cite{MiglioPop09} suggests
that these red giants are in fact stars with masses lower than 2~\msun in the low-luminosity part of the ascending RGB.
On the other hand, the same simulations concluded that the sample of red giants analysed in the CoRoT exo-field are ``Red Clump'' ones. That is, the sample is dominated by low-mass stars in the core He-burning phase. From the theoretical analysis of standard models presented in that paper, we would expect that the corresponding folded \'echelle diagram will show a significant scatter for $\ell=1$ modes.

\end {itemize}

\acknowledgements
A.M. and J.M. acknowledge financial support from the Prodex-ESA Contract Prodex 8 COROT (C90199) and FNRS.  The authors thank M.A. Dupret for useful suggestions concerning oscillation frequency computations. 


 
\end{document}